\documentclass[aps,pre,showpacs,superscriptaddress]{revtex4}

\usepackage[dvips]{graphicx}
\usepackage{mathrsfs}
\usepackage{amsmath,amssymb}
\usepackage{hhline}
\usepackage{dcolumn}
\usepackage{bm}

\usepackage[dvips]{color}

\newcommand{\rr}{\langle r \rangle}
\newcommand{\ww}{\langle w \rangle}

\newcommand{\rmax}{\langle r_{\rm max} \rangle}

\newcommand{\Rmax}{\langle R_{\rm max} \rangle}
\newcommand{\pr}{\psi}
\newcommand{\ld}{\lambda_{d}}

\newcommand{\lc}{\lambda_{c}}

\newcommand{\be}{\begin{equation}}
\newcommand{\ee}{\end{equation}}
\newcommand{\bi}{\begin{itemize}}
\newcommand{\ei}{\end{itemize}}

\newcommand{\bssa}{\beta_{\rm S}^{{\rm SW}}}
\newcommand{\bsai}{\beta_{\rm S}^{{\rm WI}}}
\newcommand{\bisa}{\beta_{\rm I}^{{\rm SW}}}
\newcommand{\biai}{\beta_{\rm I}^{{\rm WI}}}
\newcommand{\brsa}{\beta_{\rm R}^{{\rm SW}}}
\newcommand{\brai}{\beta_{\rm R}^{{\rm WI}}}
\newcommand{\basa}{\beta_{\rm W}^{{\rm SW}}}
\newcommand{\baai}{\beta_{\rm W}^{{\rm WI}}}

\begin{document}

\title{
Sudden spreading of infections in an epidemic model with a finite seed fraction
}

\author{Takehisa Hasegawa}
\email{takehisa.hasegawa.sci@vc.ibaraki.ac.jp}
\affiliation{%
Department of Mathematics and Informatics, 
Ibaraki University, 
2-1-1, Bunkyo, Mito, 310-8512, Japan
}%
\author{Koji Nemoto}
\email{nemoto@statphys.sci.hokudai.ac.jp}
\affiliation{%
Department of Physics, Hokkaido University,
Kita 10 Nishi 8, Kita-ku, Sapporo, Hokkaido, 060-0810, Japan
}%

\begin{abstract}
We study a simple case of the susceptible-weakened-infected-removed model in regular random graphs
in a situation where an epidemic starts from a finite fraction of initially infected nodes (seeds).
Previous studies have shown that, assuming a single seed, 
this model exhibits a kind of discontinuous transition at a certain value of infection rate.
Performing Monte Carlo simulations and evaluating approximate master equations, 
we find that the present model has two critical infection rates for the case with a finite seed fraction.
At the first critical rate the system shows a percolation transition of clusters composed of removed nodes, 
and at the second critical rate, which is larger than the first one, a giant cluster suddenly grows and 
the order parameter jumps even though it has been already rising.
Numerical evaluation of the master equations shows that 
such sudden epidemic spreading does occur if the degree of the underlying network is large and the seed fraction is small.
\end{abstract}


\maketitle

\section{Introduction \label{sec:introduction}}

Contagion processes such as infectious diseases, opinion formations, and information propagations, are ubiquitous in our networked society. 
Network science has shown the profound impact that network structures give on such spreading behaviors 
\cite{albert2002statistical,newman2003structure,barrat2008dynamical,dorogovtsev2008critical}.  
An epidemic can spread globally and infect a large number of individuals even if its infection rate is infinitesimally small, when the underlying network is highly heterogeneous \cite{pastor2001epidemic,moreno2002epidemic}; 
such events can be described by simple infectious disease models, such as the susceptible-infected-removed (SIR) model \cite{kermack1927contribution} and the susceptible-infected-susceptible (SIS) model \cite{anderson1992infectious}. 
Extensive studies \cite{pastor2014epidemic} on simple epidemic models in complex networks have developed accurate approximations with which the time evolution of epidemic spreading is well described \cite{miller2011edge,gleeson2011high,lindquist2011effective,gleeson2013binary}, and they have discovered new phenomena owing to the complex structure of the networks, such as the localization of epidemics \cite{goltsev2012localization,odor2014localization} and the stretching of criticality \cite{moretti2013griffiths,odor2015griffiths,cota2016griffiths}.
Alongside these studies, a similar amount of effort has been devoted to contagion processes in social dynamics 
in order to recognize the effect of network connecting individuals on human behaviors, such as opinion formations and the propagation processes of information, innovations, and transient fads \cite{castellano2009statistical}. 

Social reinforcement, which means that an individual requires multiple prompts from neighbors before adopting information, has received attention in recent years.
A generalized contagion model incorporating social reinforcement was firstly introduced by Dodds and Watts \cite{dodds2004universal,dodds2005generalized}.
Centola et al. confirmed the effect of social reinforcement in individuals' behavior through experiments on online social networks \cite{centola2007cascade,centola2010spread}.
After these seminal works, several complex contagion models have been proposed 
\cite{krapivsky2011reinforcement,zheng2013spreading,campbell2013complex,melnik2013multi,hasegawa2014discontinuous,wang2015dynamics,o2015mathematical,miller2016complex,wang2016dynamics}. 
Krapivsky et al. \cite{krapivsky2011reinforcement} put forward a mathematical model for transient fads, which we call the fad model. 
The fad model is a variant of the SIR model, in which intermediate states are added between the susceptible (S) and infected (I) states (corresponding to uninformed and adopter in the fad context, respectively).
This model was analyzed by the rate equation which assumes a well-mixed population (i.e., the mean-field approximation) and it showed that its transition is explosive in the sense that, at a certain value of infection rate, the number of individuals that adopted information suddenly grows and the order parameter discontinuously jumps from zero to a non-zero value.
Very recently, the susceptible-weakened-infected-removed (SWIR) model, which is a variation of the SIR model with an additional ``weakened'' state, has been studied \cite{lee2016universal,choi2017mixed}.
Branching process analysis and numerical simulations have demonstrated that, in random graphs, a single infected node in this model can trigger an infinite avalanche of macroscopic order; this indicates the existence of a discontinuous transition within appropriate model parameters.
We should note that the generalized epidemic process (GEP) is identical to the SWIR model and has already been studied in \cite{janssen2004generalized,bizhani2012discontinuous,chung2014generalized,janssen2016first} ahead of the SWIR studies \cite{lee2016universal,choi2017mixed}. 
In \cite{janssen2004generalized}, Janssen et al. studied the mean-field theory and renormalized field theory for the GEP in order to show that the GEP has a tricritical point of continuous and discontinuous transitions in the parameter space and to investigate its universality class in details. Bizhani et al. \cite{bizhani2012discontinuous} considered a spatial GEP (the GEP on lattices and random graphs) to show that both discontinuous and continuous transitions can occur in $d \ge 3$ dimensional lattices. A related model, namely heterogeneous $k$-core percolation, has been well studied mainly by using the local tree approximation \cite{baxter2011heterogeneous}.

In this paper, we investigate the effect of a nontrivial initial condition on the SWIR model in a network, i.e., we consider the case that an epidemic starts from a finite fraction of initially infected nodes (seeds).
Few studies have investigated the effect of such initial conditions on SIR-type epidemic models in networks \cite{miller2014epidemics,hu2014effects,ji2015effective}, although we have previously studied the SIR model with a finite seed fraction in order to show that the critical infection rate, which can be described as ``percolation'' of removed (R) nodes, crucially depends on the seed fraction $\rho$ \cite{hasegawa2016outbreaks}.
When this fraction is finite ($\rho>0$), a cluster of R nodes created by each seed grows as the infection rate increases, and before inducing a global outbreak by itself alone a number of such finite clusters connect with each other to form a giant cluster characterizing percolation of R clusters.
The critical infection rate of this SIR model is, therefore, smaller than the well known epidemic threshold.
At rate above this threshold, a single seed can spread through the network; but as long as $\rho>0$, it does not exhibit any singularities at the threshold.

The SWIR model with a single seed exhibits a discontinuous transition, while finite fractions of seeds may induce a continuous percolation transition of R nodes. 
The aim of this paper is to observe what happens when $\rho$ for the SWIR model is finite.
We perform Monte Carlo simulations for the SWIR model in a regular random graph (RRG) in order to confirm that the discontinuity of the order parameter holds for the present case, and that the model also exhibits an ordinary percolation transition at a smaller rate. 
We also investigate the SWIR model using approximate master equations (AMEs) \cite{gleeson2011high,lindquist2011effective,gleeson2013binary,hasegawa2016outbreaks} in order to find that a discontinuous transition occurs if the degree of the underlying network is large and the seed fraction is small.

\section{Model \label{sec:model}}

Let us consider the SWIR model in a static network with $N$ nodes.
Each node in the network takes one of the following four states: susceptible (S), weakened (W), infected (I), or removed (R). 
As an initial state of each trial, a fraction, $\rho$, of nodes is randomly chosen as seeds and is initially I, while other nodes are S.
The system evolves as 
\begin{eqnarray}
&&{\rm S}\; + \;{\rm I}\; \to {\rm I}\; + \;{\rm I} \quad {\rm with} \; {\rm infection} \; {\rm rate} \; \kappa,
\nonumber \\
&&{\rm S}\; + \;{\rm I}\; \to {\rm W}\; + \;{\rm I} \quad {\rm with} \; {\rm infection} \; {\rm rate} \; \lambda_1,
\nonumber \\
&&{\rm W}\; + \;{\rm I}\; \to {\rm I}\; + \;{\rm I} \quad {\rm with} \; {\rm infection} \; {\rm rate} \; \lambda_2,
\nonumber \\
&&{\rm I}\; \to {\rm R} \quad {\rm with} \; {\rm removal} \; {\rm rate} \; \mu.
\nonumber
\end{eqnarray}
Dynamics stop when I nodes no longer exist in the network. In other words, when only S, W, and R nodes remain.

The case where $\lambda_1=\lambda_2=0$ corresponds to an ordinary SIR model, whose transition is a continuous one.
The SWIR model shows a discontinuous transition when two additional processes (${\rm S} + {\rm I} \to {\rm W} + {\rm I}$ and ${\rm W} + {\rm I} \to {\rm I} + {\rm I}$) work effectively. 
Lee et al. \cite{lee2016universal} studied the time evolution of a giant cluster in an SWIR model in order to determine the mechanism causing explosive spreading to occur from a single seed.
Choi et al. \cite{choi2017mixed} investigated an SWIR model with a single seed and $\kappa=0$, $\lambda_1=\lambda$, and $\lambda_2=\mu=1$ in random graphs and found that the transition is rather mixed order; what is meant by this is that an order parameter exhibits a discontinuous jump without any critical behavior while other physical quantities exhibit critical behaviors when the average degree is large.
As already stated in Sec.~\ref{sec:introduction}, a discontinuous transition has been confirmed for more general settings in the GEP \cite{janssen2004generalized,bizhani2012discontinuous,chung2014generalized,janssen2016first}.
In our study, we consider an SWIR model with $\kappa=0$ and $\lambda_1=\lambda_2=\lambda$, which is a simplest setting for this model to exhibit a discontinuous transition \footnote{The SWIR model with $\kappa=0$ and $\lambda_1=\lambda_2=\lambda$ corresponds to the fad model containing one intermediate state, where a discontinuous transition was analytically shown in mean-field dynamics \cite{krapivsky2011reinforcement}.}.
When an S or W node is adjacent to an I node, this node becomes W or I, respectively, with probability $\lambda \Delta t$ within a short time $\Delta t$.
This probability is independently given by each of the I nodes, and so the total infection rate at an S or W node is proportional to the number of I neighbors.
An I node becomes R at a rate $\mu$ irrespective of the neighbors' states. 
Without loss of generality, we set $\mu=1$ unless otherwise specified. 
In the following sections, we focus on the RRG where all the nodes have the same degree $z$, in order to investigate dynamics in details.

In the rest of this section, we briefly review the SIR model in an RRG in order to introduce the order parameter responsible for the characterization of the phase transition.
When there is a single seed, or the fraction of infected seeds is infinitesimally small ($\rho \to 0$), the SIR model undergoes a phase transition at the epidemic threshold, $\kappa_c(\rho \to 0)$.
Let us consider the mean fraction of the outbreak size (the number of R nodes), $\rr$, where $\langle \cdot \rangle$ represents the average taken over all trials. 
In the limit $N \to \infty$, the system has two phases: (i) the local epidemic phase where $\rr=0$ for $\kappa<\kappa_c (\rho \to 0)$ and (ii) the global epidemic phase where $\rr>0$ for $\kappa>\kappa_c (\rho \to 0)$.
Around $\kappa_c$, the SIR model exhibits typical critical behaviors with the same universality class as (dynamic) percolation \cite{tome2010critical, de2011new}.

When the dynamics start from a finite fraction of I nodes, $\rho>0$, the phase transition can be described in terms of percolation \cite{stauffer1992introduction}.
Connected components of R nodes are hereafter called R clusters. 
In the case of $\rho >0$, a final state will contain numerous R clusters.
The order parameter can then be given as $\rmax=\Rmax/N$, where $\Rmax$ is the mean size of the largest R cluster.
The SIR model has a critical infection rate $\kappa_c(\rho)$ such that $\rmax$ decreases to zero as $N$ increases for $\kappa<\kappa_c(\rho)$ and converges to a nonzero value with increasing $N$ for $\kappa>\kappa_c(\rho)$. 
The phase transition at $\kappa_c(\rho)$ can be described in terms of percolation and its critical exponents are the same as those of percolation in the same network \cite{hasegawa2016outbreaks}.
The critical infection rate $\kappa_c(\rho)$ depends on the seed fraction $\rho$, and deviates rapidly from the epidemic threshold $\kappa_c(\rho \to 0)$ as $\kappa_c(\rho \to 0)-\kappa_c(\rho) \sim \rho^{1/3}$ \cite{hasegawa2016outbreaks}.
Note that $\rr$ is not appropriate to describe the (continuous) transition for $\rho > 0$ because $\rr > \rho >0$, while, in the single seed case, R cluster is always unique and thus $\rr=\rmax$. 

The above will hold for the SWIR model. If a continuous percolation transition of R nodes is induced by a finite fraction of seeds, its transition will be characterized by $\rmax$, but not by $\rr$. On the other hand, when a discontinuous transition of epidemic spreading occurs, it will be reflected both on $\rmax$ and $\rr$. In the Sec.~\ref{sec:result1} and Sec.~\ref{sec:result2}, we calculate $\rmax$ using Monte Carlo simulations and $\rr$ using the approximate master equations, respectively, to confirm the coexistence of continuous percolation transition and discontinuous epidemic spreading in the SWIR model.

\section{Monte Carlo simulations \label{sec:result1}}

\begin{figure}
 \begin{center}
  \includegraphics[width=75mm]{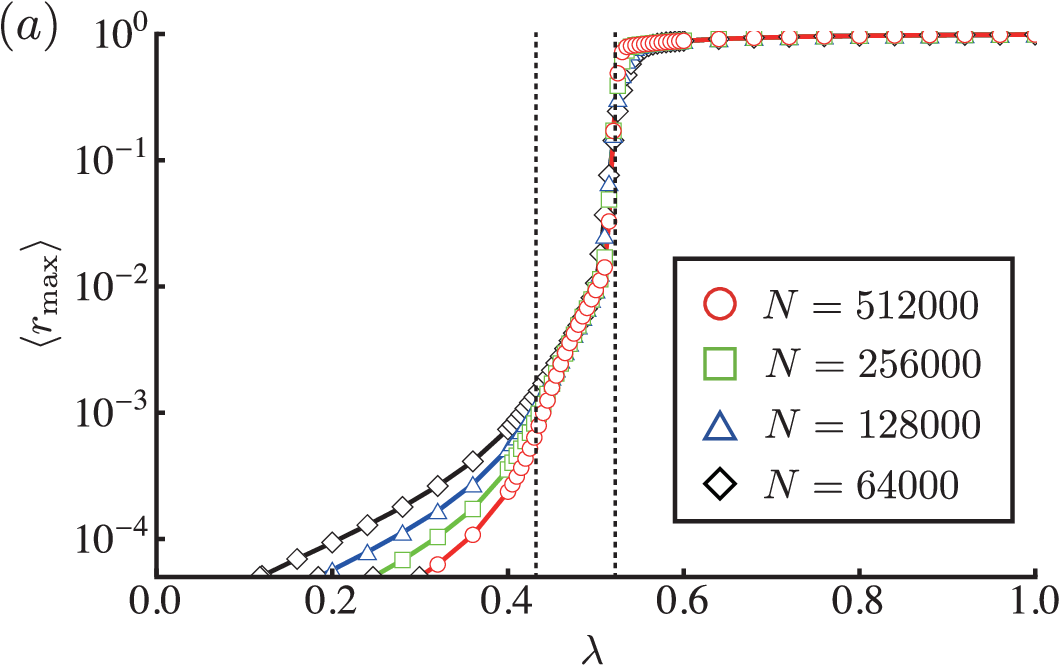}
  \includegraphics[width=75mm]{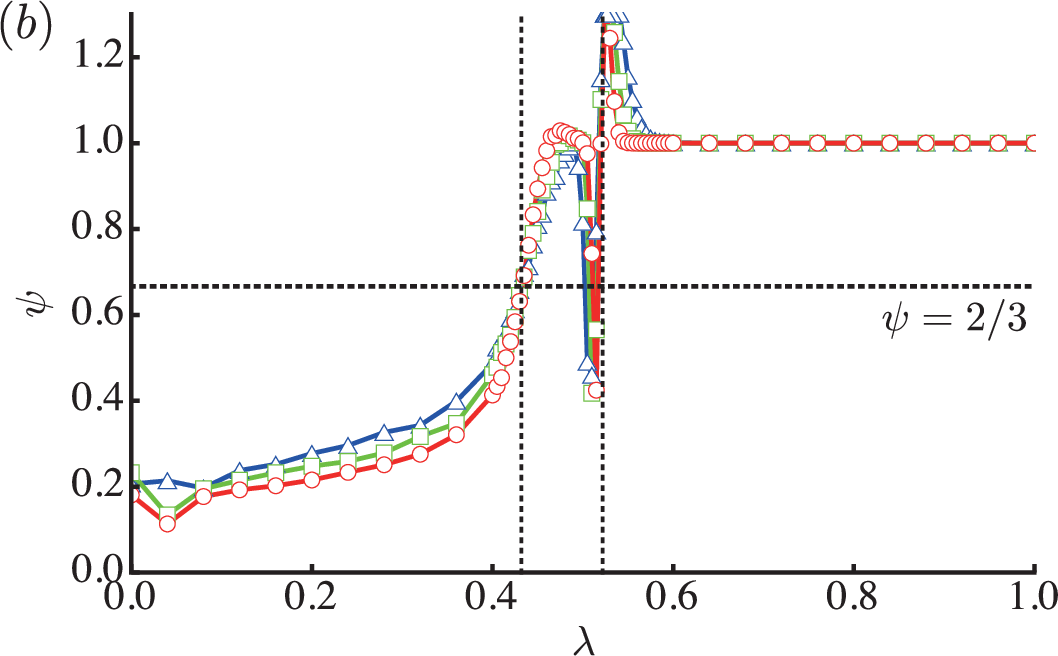}
   \end{center}
 \caption{
Numerical results of (a) the order parameter $\rmax(N)$ and (b) the fractal exponent $\psi(N)$ for the SWIR model in the RRG with $z=9$.
The two dotted lines represent $\lc \approx 0.432$ (left) and $\ld \approx 0.52$ (right).
}
 \label{fig-orderparameter}
\end{figure}

First, we perform Monte Carlo simulations for the SWIR model in the RRG with $z=9$.
The number of nodes is $N=64000, 128000, 256000$, and $512000$. 
The number of graph realizations is 100, and the number of trials on each graph is 500.
Seeds are randomly chosen at each trial and the fraction of seeds is $\rho=0.001$.

The order parameter $\rmax$ is shown in Fig.~\ref{fig-orderparameter}~(a) for the RRGs of several sizes.
The $N$-dependence of the order parameter indicates that there are two critical infection rates: $\lc$, below which $\rmax$ tends to zero as $N$ increases and above which $\rmax$ converges to a nonzero value; and $\ld$, around which $\rmax$ rises more sharply in larger networks and eventually shows a discontinuous jump in the limit $N \to \infty$. 

In order to obtain the first critical rate, $\lc$, numerically, we evaluate the fractal exponent, $\pr$ ($0 \le \pr \le 1$), of the largest R cluster, defined as $\Rmax \propto N^\pr$ \cite{hasegawa2013profile,hasegawa2014critical}.
The fractal exponent in finite networks, $\pr(N)$, is given and approximated by $\pr(N)={\rm d} \ln \Rmax(N)/{\rm d} \ln N \approx (\ln \Rmax(N) - \ln \Rmax(N/2))/(\ln N - \ln (N/2))$.
As $N$ increases, $\pr(N)$ goes to zero for $\lambda<\lc$ and to one for $\lambda>\lc$, thereby reflecting that $\Rmax \sim O(1)$ for $\lambda<\lc$ and $\sim O(N)$ for $\lambda>\lc$.
As shown in Fig.~\ref{fig-orderparameter}~(b), $\pr(N)$ of several networks of different sizes crosses at a unique point, giving $\lc \approx 0.432$. 
The transition at $\lc$ is of an ordinary percolation, $\pr \approx 2/3$ at $\lambda=\lc$, and $\Rmax$ around $\lc$ obeys a finite size scaling \cite{hasegawa2013profile} if $\rmax(\lambda)-\rmax(\lc) \sim |\lambda-\lc|^\beta$ with $\beta=1$ is assumed (not shown). 
The values of $\pr$ and $\beta$ obtained here are also observed for the SIR model with $\rho>0$ in the RRG \cite{hasegawa2016outbreaks}.  
An apparent oscillatory deviation from $\psi=1$ is seen at  $\ld \approx 0.52$; however, this deviation is a finite size effect due to the occurrence of the discontinuous jump.
Indeed, the width of  the deviating region tends to shrink as the size increases.

\begin{figure}
 \begin{center}
  \includegraphics[width=55mm]{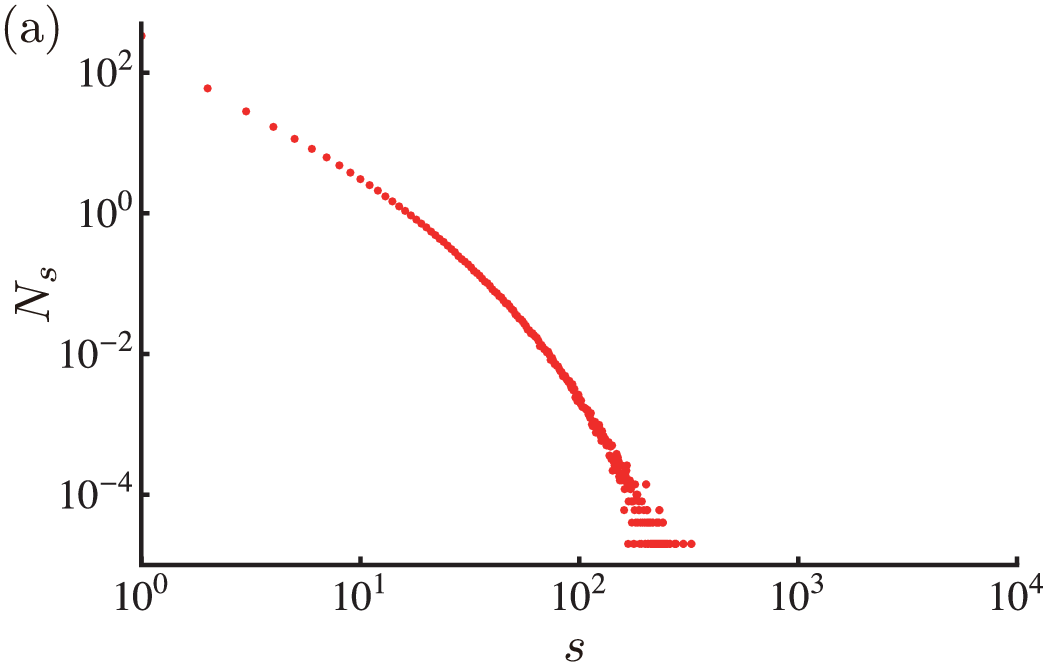}
  \includegraphics[width=55mm]{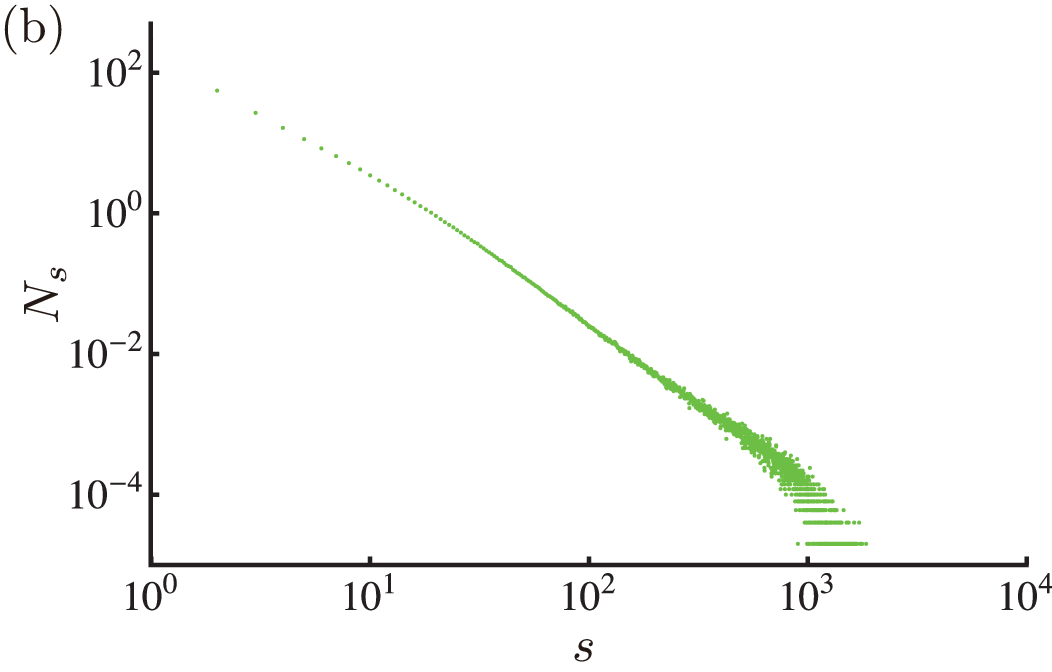}
  \includegraphics[width=55mm]{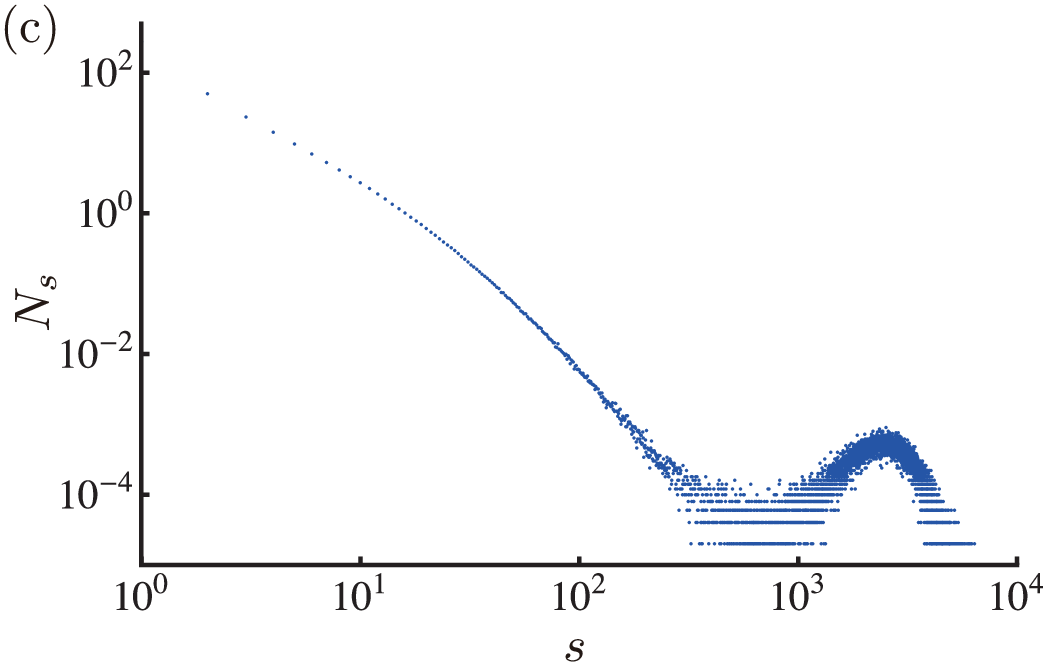}
  \includegraphics[width=55mm]{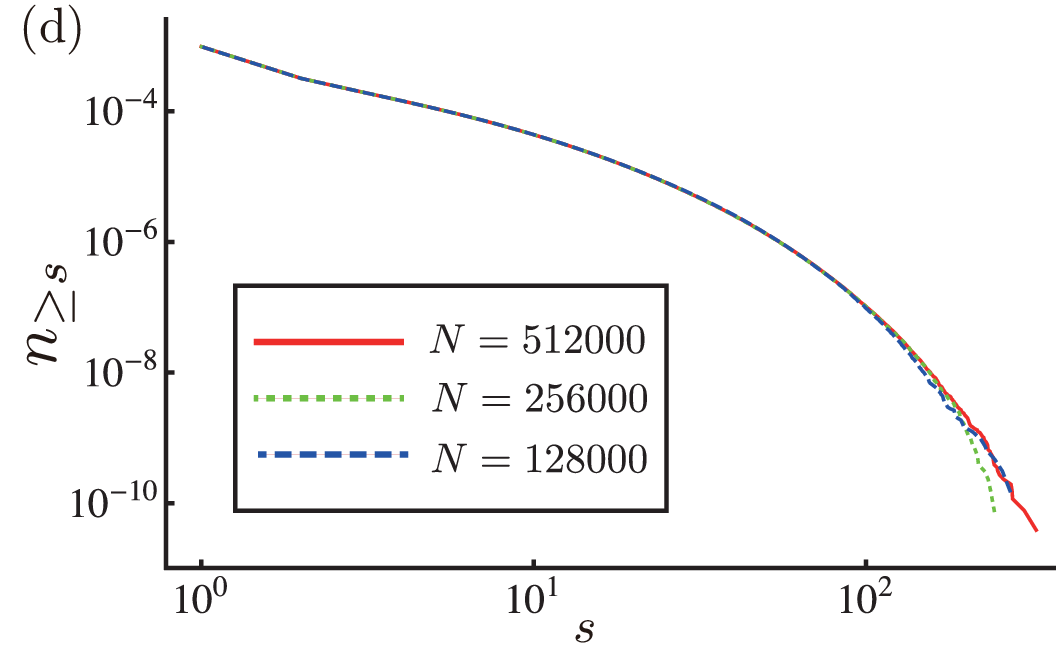}
  \includegraphics[width=55mm]{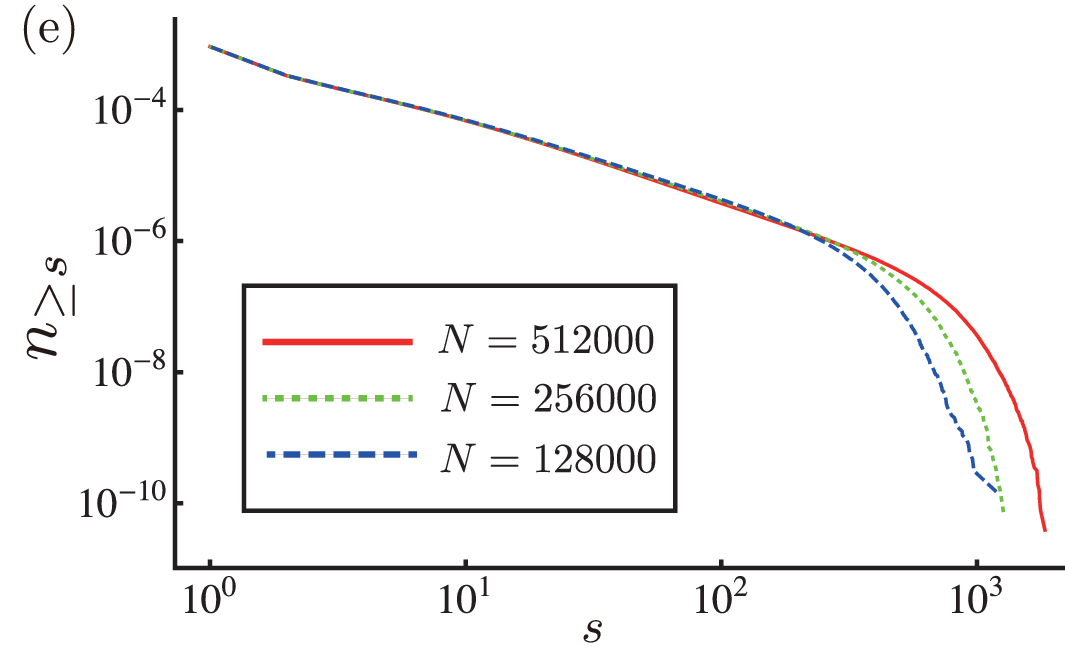}
  \includegraphics[width=55mm]{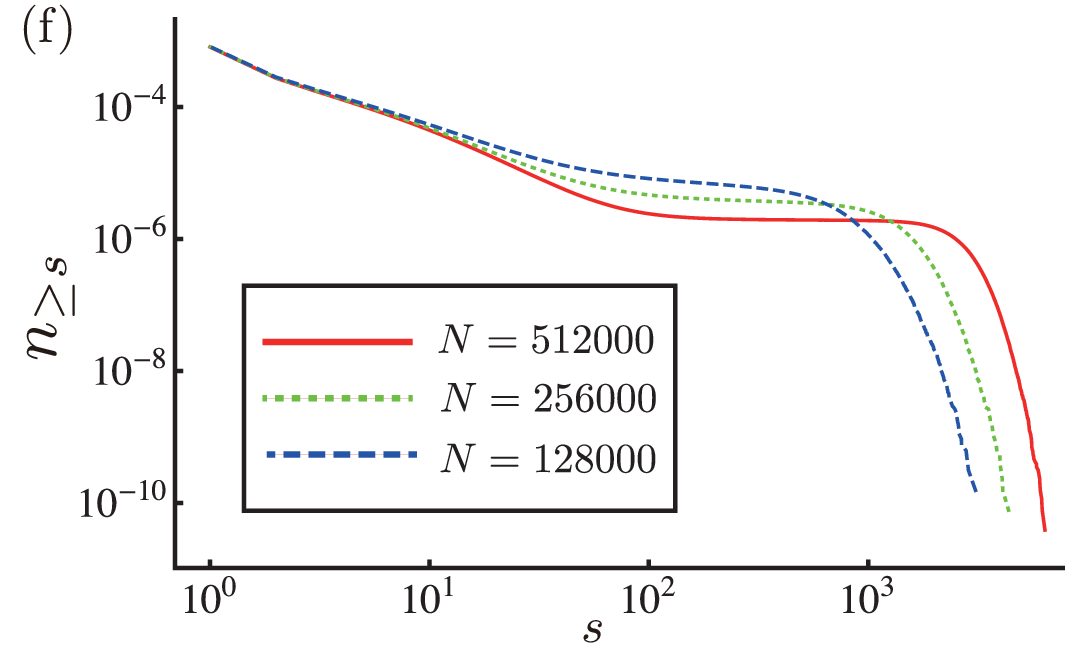}
 \end{center}
 \caption{
Mean number, $N_s$, of R clusters with size $s$ of the SWIR model in the RRG with $N=512000$, for (a) $\lambda=0.36 <\lc$, (b) $\lambda=0.43 \approx \lc$, and (c) $\lambda=0.48>\lc$. 
The average is taken over $500 \times 100$ trials. 
The cumulative distribution $n_{\ge s} = \sum_{s' \ge s} N_{s'}/N$ with several values of $N$ for (d) $\lambda=0.36 <\lc$, (e) $\lambda=0.43 \approx \lc$, and (f) $\lambda=0.48>\lc$.
 }
 \label{fig-clusterDistribution}
\end{figure}

A giant R cluster that occupies a finite fraction of the network emerges at $\lc \approx 0.432$, but its fraction remains very small (at most $10^{-2}$) as long as $\lambda<\ld$. 
To confirm the emergence of a giant cluster at $\lc$, we plot the mean number, $N_s$, of R clusters with size $s$ around $\lambda_c$ (Figs.~\ref{fig-clusterDistribution}(a)--(c)). 
We find that $N_s$ follows a typical behavior of an ordinary percolation transition; $N_s$ decays exponentially with $s$ for $\lambda<\lc$ (Fig.~\ref{fig-clusterDistribution} (a)), obeys a power-law at $\lambda=\lc$ (Fig.~\ref{fig-clusterDistribution} (b)), and becomes a bimodal distribution for $\lambda>\lc$ (Fig.~\ref{fig-clusterDistribution} (c)), respectively. 
The corresponding cumulative distributions, defined as $n_{\ge s} = \sum_{s' \ge s}N_{s'}/N$, are plotted in Figs.~\ref{fig-clusterDistribution} (d)--(f).
As shown in Fig.~\ref{fig-clusterDistribution} (d) and Fig.~\ref{fig-clusterDistribution} (e), 
the exponential and power-law decays are retained irrespective of $N$.
And as shown in Fig.~\ref{fig-clusterDistribution} (f), $N_s$ becomes bimodal more clearly in larger networks for $\lambda>\lc$, confirming the emergence of a giant cluster at $\lc$.

\begin{figure}
 \begin{center}
  \includegraphics[width=75mm]{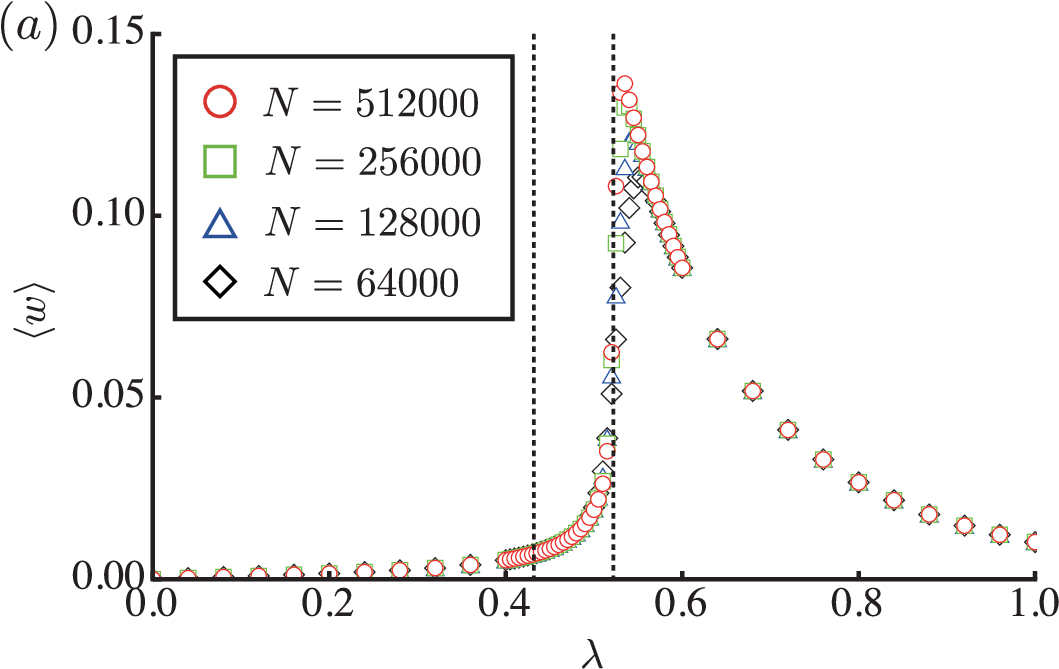}
  \includegraphics[width=75mm]{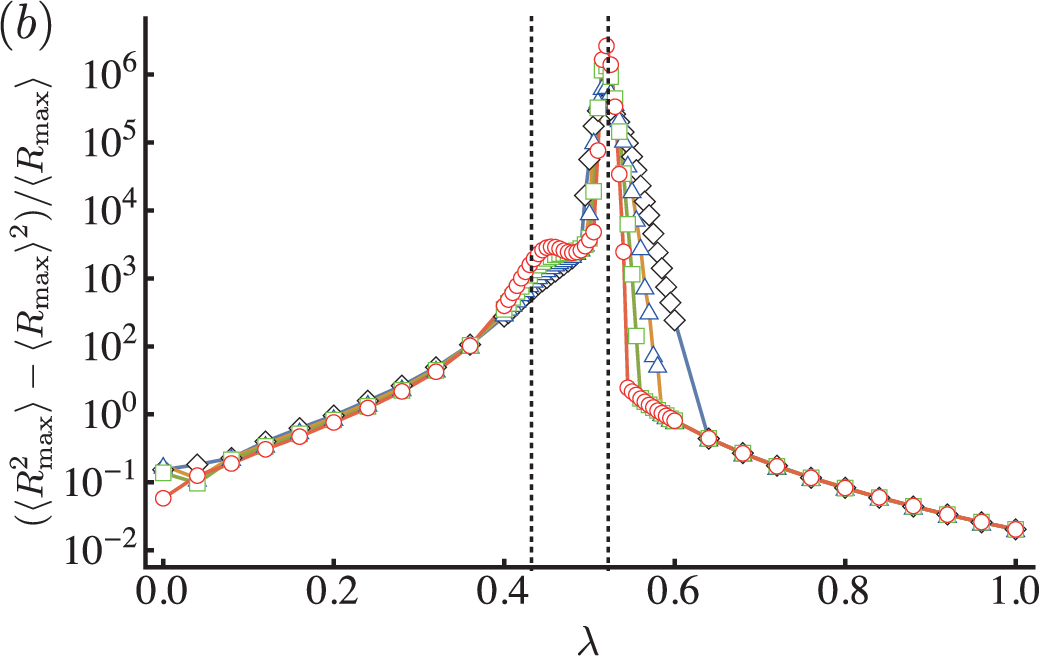}
   \end{center}
 \caption{
(a) The mean fraction of W nodes $\ww$ and (b) the sample-to-sample fluctuation of the largest cluster size, as a function of $\lambda$.
The two dotted lines represent $\lc \approx 0.432$ (left) and $\ld \approx 0.52$ (right).
}
 \label{fig-S1}
\end{figure}

\begin{figure}
 \begin{center}
  \includegraphics[width=55mm]{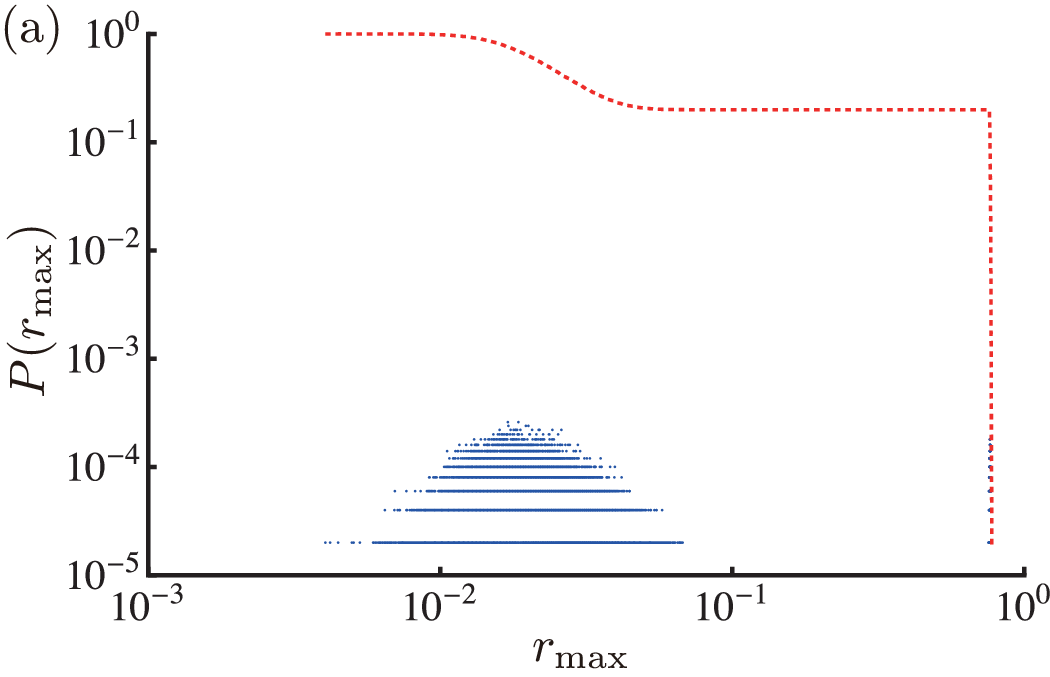}
  \includegraphics[width=55mm]{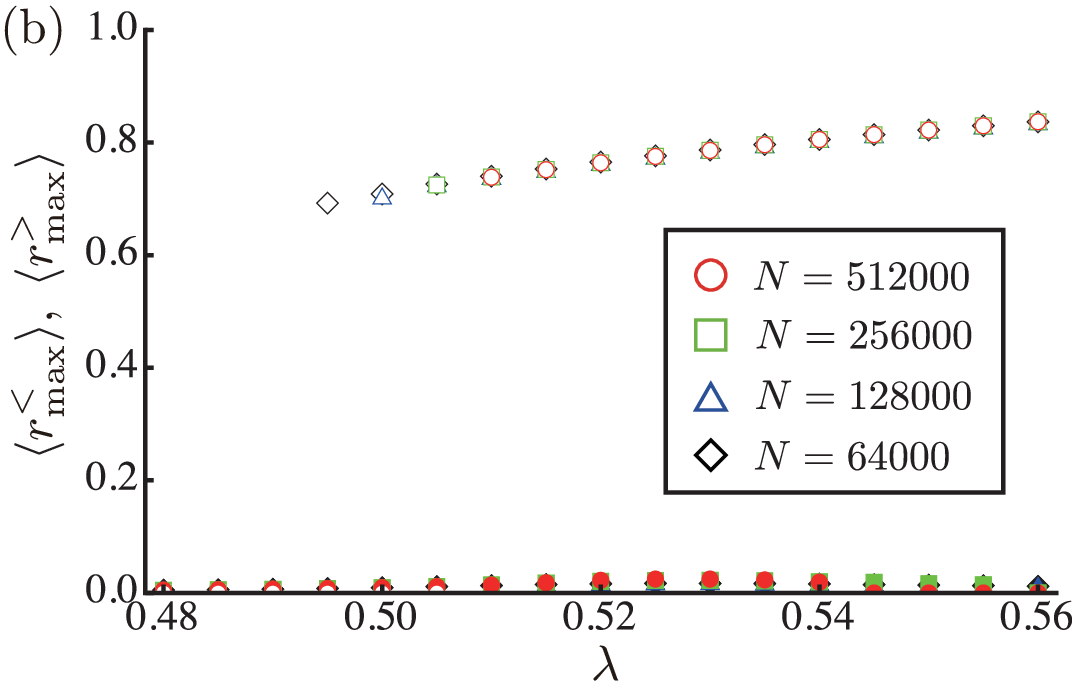}
  \includegraphics[width=55mm]{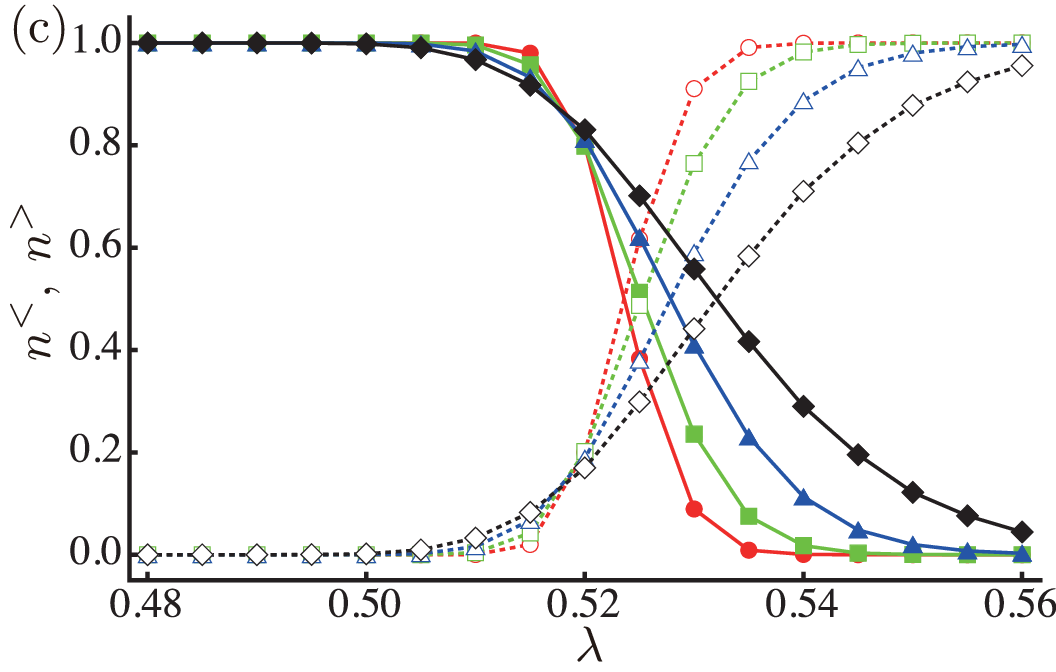}
 \end{center}
 \caption{
 (a) Distribution, $P(r_{\rm max})$, of the largest cluster fraction $r_{\rm max}$ at $\lambda = 0.52$. The red-dotted line represents the cumulative distribution $\sum_{r_{\rm max}' \ge r_{\rm max}} P(r_{\rm max}')$. 
 (b) $\langle r_{\rm max}^< \rangle$ (the full symbols) and $\langle r_{\rm max}^> \rangle$ (the open symbols), which are the largest cluster fraction averaged over samples such that $r_{\rm max}<0.5$ and $r_{\rm max}>0.5$, respectively. 
 (c) $n^<$ (the full symbols) and $n^>$ (the open symbols), which are the fractions of samples such that $r_{\rm max}<0.5$ and $r_{\rm max}>0.5$, respectively. 
 }
 \label{fig-RmaxDistribution}
\end{figure}

At $\lambda_d \approx 0.52$, the order parameter jumps from $\rmax \approx 10^{-2}$ to $\rmax \approx 0.7$. This transition is characteristic to the SWIR model and is due to the additional state W, as has been reported in previous studies \cite{lee2016universal}.
As shown in Fig.~\ref{fig-S1}~(a), the fraction of W nodes $\ww$ shows characteristic behavior around $\ld$: $\ww$ increases with $\lambda$ below $\ld$, jumps to a peak value at $\ld$, and then decreases above $\ld$.
A large number of W nodes are necessary for the strong activation of the ${\rm W}+{\rm I} \to {\rm I}+{\rm I}$ process to occur just before the explosive growth of R clusters, as has previously been mentioned for the case of single seed \cite{lee2016universal}.
This also holds for the present case.
In the case of a finite seed fraction, such abundant W nodes are supplied more efficiently because a number of W nodes peripheral to the finite R clusters are created by abundant seeds.

Although the largest R cluster is already of macroscopic order for $\lambda>\lc$, it remains very small for $\lc<\lambda<\ld$ and grows explosively at $\ld$.
Around $\ld$, the largest cluster may be created by an explosive epidemic spreading or else by a percolation of R nodes. 
Figure~\ref{fig-RmaxDistribution} (a) plots a distribution, $P(r_{\rm max})$, of the largest cluster fraction $r_{\rm max}$ near $\ld$.
The distribution $P(r_{\rm max})$ is bimodal whose peaks are located around $r_{\rm max} \approx 10^{-2} (>0)$ and $r_{\rm max} \approx 0.7$. 
The largest cluster of each sample belongs to either of two peaks; only samples such that explosive spreadings occur contribute to the larger peak.
Figure~\ref{fig-RmaxDistribution} (b) shows $\langle r_{\rm max}^< \rangle$ and $\langle r_{\rm max}^> \rangle$ characterizing the respective peak positions, which are the largest cluster fractions averaged over samples such that $r_{\rm max}<0.5$ and $r_{\rm max}>0.5$, respectively. 
We find that these averages are almost independent of $N$, while the rate at which the upper branch $\langle r_{\rm max}^> \rangle$ emerges depends on $N$. 
In Fig.~\ref{fig-RmaxDistribution} (c), we also plot $n^<$ and $n^>$, which are the fractions of samples such that $r_{\rm max}<0.5$ and $r_{\rm max}>0.5$, respectively.
We find that $n^>$ ($n^<$) increases (decreases) more sharply for larger networks.
We collect infection rates at which $n^>(N)=n^<(N)=1/2$ and extrapolate the collected data to $N \to \infty$ to obtain a rough estimate of $\ld \approx 0.52$, which is consistent with other estimates. 
An indication of two transitions is also observed for the sample-to-sample fluctuation of the largest R cluster size, $(\langle R_{\rm max}^2 \rangle-\Rmax^2)/\Rmax$, as shown in Fig.~\ref{fig-S1}~(b). 
We find that $(\langle R_{\rm max}^2 \rangle-\Rmax^2)/\Rmax$ shows a higher peak at $\ld$ than at $\lc$, although both peaks grow with $N$ and eventually diverge when $N \gg 1$. 
Our numerical results so far mean that the present model has two transitions; a continuous transition of percolation of R nodes at $\lc$ and a discontinuous transition of an explosive epidemic spreading at $\ld$.

\section{approximate master equations \label{sec:result2}}

Next, we describe the SWIR model in an infinitely large RRG of $z \ge 2$ using the approximate master equations (AMEs) \cite{gleeson2011high,lindquist2011effective,gleeson2013binary,hasegawa2016outbreaks}.
Let $s_{\ell, m, n}(t)$, $w_{\ell, m, n}(t)$, $i_{\ell, m, n}(t)$, and $r_{\ell, m, n}(t)$ be the fractions of S, W, I, and R nodes, respectively, having $\ell$ susceptible, $m$ weakened, and $n$ infected neighbors (the remaining $\tilde{z}=z-\ell-m-n$ neighbors are removed) at time $t$.
The conservation law, $s_{\ell, m, n}(t)+w_{\ell, m, n}(t)+i_{\ell, m, n}(t)+r_{\ell, m, n}(t)=1$, holds for any value of $t$.
Note that S and W nodes change their states to W and I, respectively, with the rate $\lambda n$ if they have $n$ infected neighbors.

\begin{figure}
 \begin{center}
  \includegraphics[width=75mm]{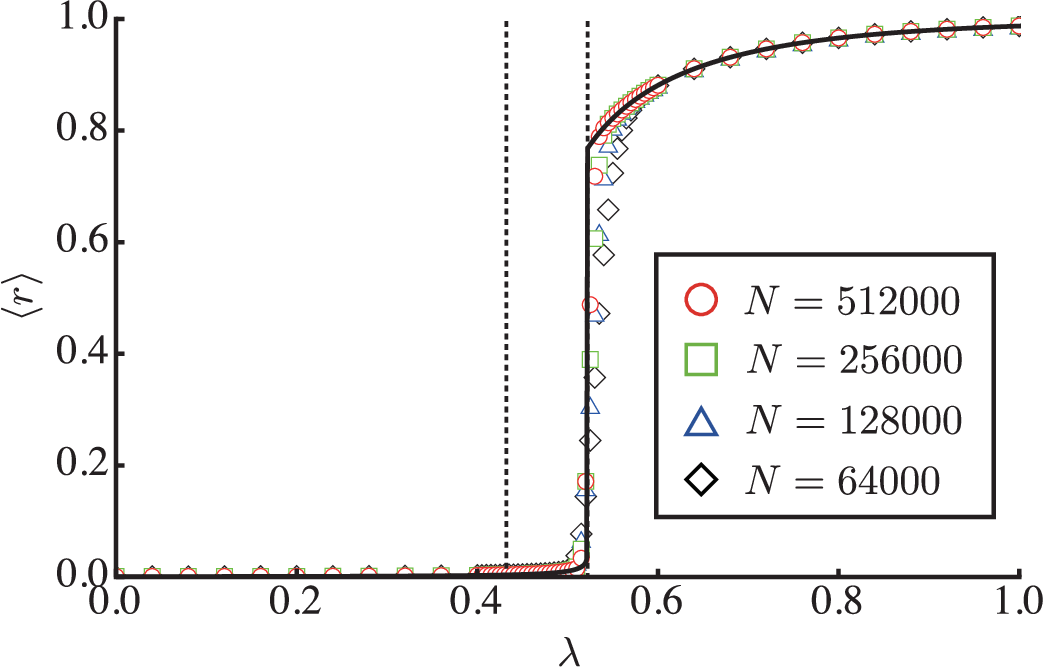}
   \end{center}
 \caption{
The fraction of removed nodes, $\rr$, as a function of $\lambda$ for the SWIR model in the RRG with $z=9$.
The solid line is drawn from the AMEs, and the symbols represent Monte Carlo results.
The two dotted lines represent $\lc \approx 0.432$ (left) and $\ld \approx 0.52$ (right).
}
 \label{fig:AME-r}
\end{figure}

Following \cite{hasegawa2016outbreaks} (see also \cite{lindquist2011effective,gleeson2013binary}), we obtain the master equation for the evolution of each density as 
\begin{subequations}
\begin{align}
\dot s_{\ell, m, n} =& 
-\lambda n s_{\ell, m, n} + \bssa [(\ell+1)s_{\ell+1,m-1,n}-\ell s_{\ell, m, n}] 
\nonumber \\ 
&+ \bsai [(m+1)s_{\ell,m+1,n-1}-m s_{\ell, m, n}]+ \mu [(n+1)s_{\ell,m,n+1}- n s_{\ell, m, n}], \label{eq:ames-s}
\\
\dot w_{\ell, m, n} =& 
\lambda n s_{\ell, m, n}- \lambda n w_{\ell, m, n} + \basa [(\ell+1)w_{\ell+1,m-1,n}-\ell w_{\ell, m, n}] 
\nonumber \\
&+ \baai [(m+1)w_{\ell,m+1,n-1}-m w_{\ell, m, n}]+ \mu [(n+1)w_{\ell,m,n+1}- n w_{\ell, m, n}], \label{eq:ames-w}
\\
\dot i_{\ell, m, n} =&
\lambda n w_{\ell, m, n} - \mu i_{\ell, m, n}+ \bisa [(\ell+1)i_{\ell+1,m-1,n}-\ell i_{\ell, m, n}]
\nonumber \\
&+ \biai [(m+1)i_{\ell,m+1,n-1}-m i_{\ell, m, n}]+ \mu [(n+1)i_{\ell,m,n+1}- n i_{\ell, m, n}], \label{eq:ames-i}
\\
\dot r_{\ell, m, n} =& 
\mu i_{\ell, m, n}+ \brsa [(\ell+1)r_{\ell+1,m-1,n}-\ell r_{\ell, m, n}]
\nonumber \\
&+ \brai [(m+1)r_{\ell,m+1,n-1}-m r_{\ell, m, n}]+ \mu [(n+1)r_{\ell,m,n+1}- n r_{\ell, m, n}]. \label{eq:ames-r}
\end{align}
\label{eq:ame}
\end{subequations}
In each of these equations, $\beta_{\rm X}^{\rm YZ}$ represents the rate at which a randomly chosen neighbor of a randomly chosen node in the state X ($=$ S, W, I, R) changes its state from Y ($=$ S, W) to Z ($=$ W, I); this can be approximated as the rate at which an X-Y edge is changed to an X-Z edge \cite{gleeson2013binary}, as shown in the below equations.
\begin{subequations}
\begin{align}
&
\bssa=\lambda\frac{\sum_{\ell, m, n} \ell n s_{\ell, m, n}}{\sum_{\ell, m, n}\ell s_{\ell, m, n}}, \; 
\basa=\lambda\frac{\sum_{\ell, m, n} m n s_{\ell, m, n}}{\sum_{\ell, m, n}m s_{\ell, m, n}}, \; 
\bisa=\lambda\frac{\sum_{\ell, m, n} n^2 s_{\ell, m, n}}{\sum_{\ell, m, n}n s_{\ell, m, n}}, \; 
\brsa=\lambda\frac{\sum_{\ell, m, n} \tilde{z} n s_{\ell, m, n}}{\sum_{\ell, m, n} \tilde{z} s_{\ell, m, n}}, \\
&
\bsai=\lambda\frac{\sum_{\ell, m, n} \ell n w_{\ell, m, n}}{\sum_{\ell, m, n} \ell w_{\ell, m, n}}, \;
\baai=\lambda\frac{\sum_{\ell, m, n} m n w_{\ell, m, n}}{\sum_{\ell, m, n}m w_{\ell, m, n}}, \;
\biai=\lambda\frac{\sum_{\ell, m, n} n^2 w_{\ell, m, n}}{\sum_{\ell, m, n}n w_{\ell, m, n}}, \;
\brai=\lambda\frac{\sum_{\ell, m, n} \tilde{z} n w_{\ell, m, n}}{\sum_{\ell, m, n} \tilde{z} w_{\ell, m, n}}. 
\end{align}
\end{subequations}
In these equations, the summations run over all $0 \le \ell+m+n \le z$.
The initial state where a fraction $\rho ( > 0 )$ of nodes is randomly chosen as seeds is given by
\begin{equation}
s_{\ell,m,n}(0)=\delta_{z,\ell+n}(1-\rho){z \choose \ell}(1-\rho)^\ell \rho^n, \quad
i_{\ell,m,n}(0)=\delta_{z,\ell+n}\rho{z \choose \ell}(1-\rho)^\ell \rho^n, \quad
w_{\ell,m,n}(0)=r_{\ell,m,n}(0)=0. \label{eq:initCond}
\end{equation}
By evaluating the AMEs (\ref{eq:ame}) with this initial condition, we obtain the total density of each state at time $t$ as
\begin{equation}
s(t)=\sum_{\ell,m,n}s_{\ell,m,n}(t),\quad
w(t)=\sum_{\ell,m,n}w_{\ell,m,n}(t),\quad
i(t)=\sum_{\ell,m,n}i_{\ell,m,n}(t),\quad
r(t)=\sum_{\ell,m,n} r_{\ell,m,n}(t). \label{eq:AMEtotal}
\end{equation}
In Fig.~\ref{fig:AME-r}, we plot the fraction of R nodes, $\rr$, in the final states using the AMEs and the Monte Carlo simulations.
Here $\rr = r(t \to \infty)$ for the AMEs.
The AMEs clearly show that there is a discontinuous jump of $\rr$ at $\ld \approx 0.52$, where a jump of $\rmax$ has already been observed numerically.
Indeed, the Monte Carlo simulations support the solutions of the AMEs: data related to $\rr(N)$ approaches the line drawn by the AMEs as $N$ increases. 

\begin{figure}
 \begin{center}
  \includegraphics[width=75mm]{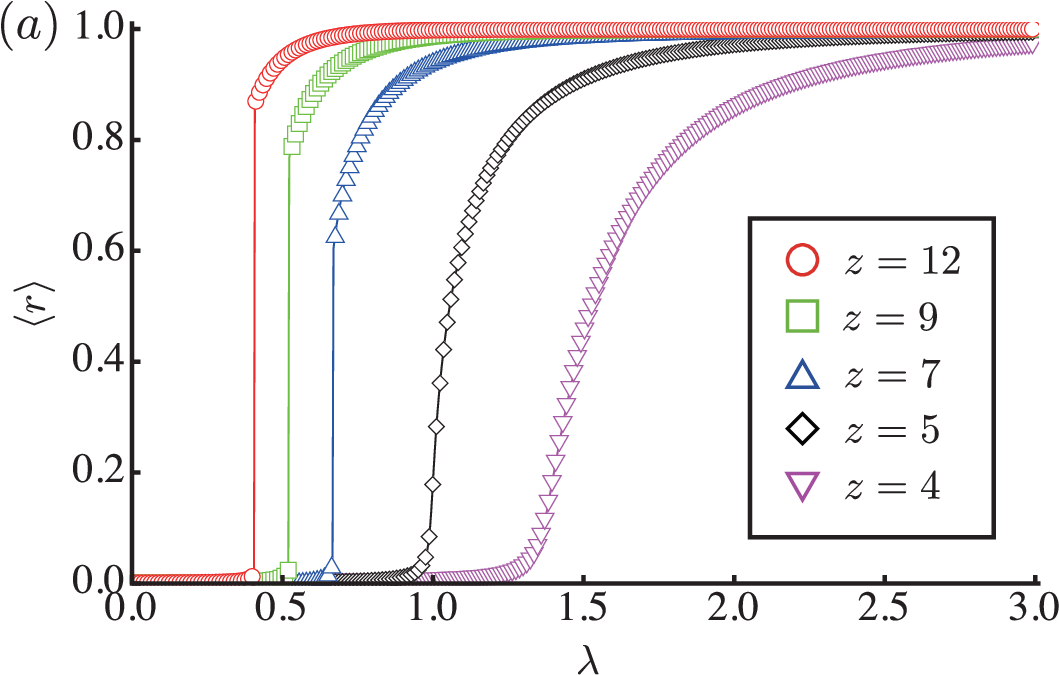}
  \includegraphics[width=75mm]{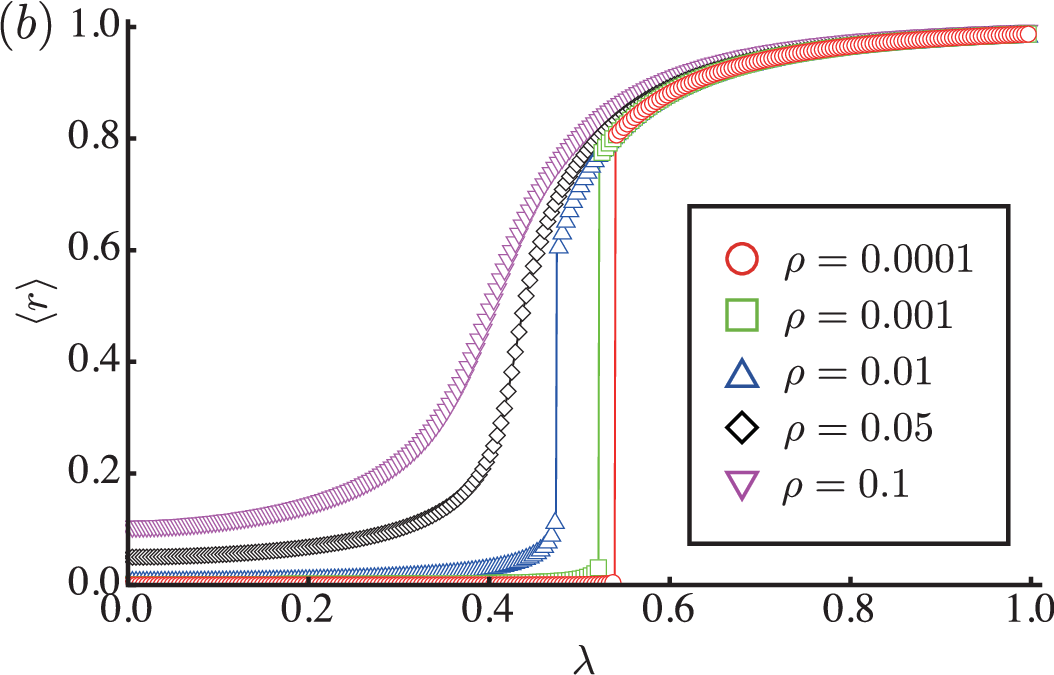}
   \end{center}
 \caption{
(a) Mean final fraction of the removed nodes, $\rr$, in the RRGs with $\rho=0.001$ and $z = 12, 9, 7, 5$, and $4$ (from the left to the right), and (b) mean final fraction of the removed nodes, $\rr$, in the RRG with $z=9$, and $\rho=0.1,0.05,0.01,0.001$, and $0.0001$ (from the left to the right).
The symbols and lines are obtained by evaluating the AMEs.
}
 \label{fig:AME-dependence}
\end{figure}

The discontinuous transition at $\ld$ is reflected both on $\rmax$ and $\rr$.
Utilizing the AMEs for the SWIR model in the RRG with given values of $\rho$ and $z$, we obtain the final density of each state with high accuracy and a short computation time, considering the $z$-dependence and $\rho$-dependence of $\rr$. 
In Fig.~\ref{fig:AME-dependence}~(a), we plot $\rr$ for the case of $\rho=0.001$ when the degree of the RRG changes ($z=12,9,7,5$, and $4$).
For a fixed $\rho$, the discontinuity of $\rr$ is clearer in networks with larger degrees. 
This discontinuity, however, becomes weak as $z$ decreases and it disappears in RRGs with small degrees of $z$ (i.e., $z \le 5$).
In Fig.~\ref{fig:AME-dependence}~(b), we plot $\rr$ for the case where $z=9$ when the seed fraction changes ($\rho=0.1,0.05,0.01,0.0001$, and $0.0001$).
We find that there is a discontinuous jump of $\rr$ when $\rho$ is small; it becomes less prominent as $\rho$ increases, however, and it is eventually hard to distinguish an explosive spreading from the percolation of R clusters.
We observe that the phase transition of the SWIR model with multiple seeds consists of two components: a percolation of R clusters and an explosive spreading of infections. 
The latter, however, disappears when $\rho$ is too large and/or $z$ is too small.

\section{Summary \label{sec:summary}}

In this paper, we studied the spreading behaviors of the SWIR model in the RRGs. 
As previous studies have shown, the SWIR model starting from a single infected node exhibits a discontinuous transition at a certain infection rate, but when starting from a finite fraction of infected seeds, the model shows three different regimes that depends on the infection rate: the local epidemic phase, $\lambda<\lambda_c$, where clusters of removed nodes remains finite; the low global epidemic phase, $\lambda_c < \lambda < \lambda_d$, where a giant cluster of removed nodes (i.e., a giant R cluster) exists but the total fraction of removed nodes remains small; and the high global epidemic phase, $\lambda > \lambda_d$, where a giant R cluster occupies a large part of the network. 
We showed numerically that the first transition at $\lambda_c$ is continuous in the same manner as an ordinary percolation transition, and that the second transition at $\lambda_d$ is discontinuous.
We formulated AMEs for the present model in order to demonstrate that the phase transition critically depends on both the degree of the underlying network and the fraction of initial seeds: the discontinuous transition disappears when the degree is small or when the seed fraction is large.
The coexistence of continuous transition (at $\lc$) and discontinuous transition (at $\ld$) is characteristic of the SWIR model with multiple seeds.
If there is no additional state W of the SWIR model (the SIR model), a discontinuity at $\ld$ is not observed and there is only continuous transition of percolation \cite{hasegawa2016outbreaks}.
If the SWIR epidemic starts with a single seed or infinitesimal seed fraction, the percolation of R nodes is absorbed by the discontinuous transition at $\ld$.

We focused on a simple version of the original SWIR model ($\kappa=0$, $\lambda_1=\lambda_2=\lambda$, and $\mu=1$). 
However, the present results are expected to hold true for the SWIR model itself. 
In \cite{hasegawa2014discontinuous}, we have studied a multi-stage independent cascade model, which corresponds to a discrete-time SWIR model of $\lambda_2=\kappa+\lambda_1$, and observed the same behavior, i.e., a discontinuous change of the order parameter in a percolating region. 
Although the SWIR model and the GEP with finite seed fractions have been previously examined
\cite{janssen2004generalized,bizhani2012discontinuous,chung2014generalized,janssen2016first,choi2017critical}, those studies have focused only on the discontinuous transition at $\ld$ and missed the continuous percolation transition at $\lc$. 
The reason may be that the mean-field and local tree approximations mainly used in previous studies fail to predict the percolation transition at $\lc$. 
The mean-field theory of a system is equivalent to consider a model on the complete graph with taking normalized infection rate. 
As each node connects with all other nodes in the complete graph, the order parameter $r_{\rm max}$ is always equal to the total density of R nodes, $r_{\rm max}=r$. 
Then, if $\rho > 0$, the mean-field theory just says that R nodes always percolates irrespective of $\lambda$ because $r_{\rm max} = r \ge \rho > 0$. 
In other words, $\lambda_c=0$ as long as $\rho>0$. 
As to the local tree approximation, Choi et al. \cite{choi2017critical} reported a discontinuous transition for the SWIR model of multiple seeds. However, they considered the total density of R nodes as an order parameter and did not calculate the percolation transition at $\lc$. 
More to say, the local tree approximation in \cite{choi2017critical} assumes that an R node and neighboring R nodes in a cluster share at least one edge through which an infection event occurred. 
However, even if not so, two R nodes become members of the same cluster when ``two R nodes are adjacent'', and there can be a giant cluster. Such a giant cluster will not be detected by self-consistent formalism. 
These circumstances require us to calculate the transitions for the complex contagion models with nontrivial initial conditions in details.

As mentioned in Sec.~\ref{sec:introduction}, the present model is also used as a propagation model for innovations and transient fads \cite{krapivsky2011reinforcement}. 
The seed fraction determining whether a transition is continuous or discontinuous may be an important issue for viral marketing campaigns. 
If there are few initial adopters (seeds), a viral marketing campaign becomes an all-or-nothing venture, i.e., there is only either complete success or complete failure, up to the influence between individuals.
The cost effectiveness of marketing is otherwise rather reasonable: increasing initial adopters, or making the information more attractive, is directly linked to the number of individuals who will adopt the information. 

Infectious diseases with finite fractions of infected seeds have been studied less than cases where there are single seed or infinitesimally small seed fractions. 
We know of almost no other complex epidemic models with finite seed fractions.
Among complex epidemic models with finite seed fractions, a prospective one is cooperative epidemics \cite{chen2013outbreaks,cai2015avalanche,hebert2015complex,grassberger2016phase,azimi2016cooperative}; in these, the infection of one disease stimulates the spreading of others. 
It will be interesting to see exactly how, and if, numerous seeds could cause catastrophic co-infections.
In this study, we did not study the effect of disorder on the present model. As is known, topological disorder of a network and quenched disorder on infection rates of nodes can induce the Griffiths phase (the stretching of criticality) for the SIS model \cite{juhasz2012rare}. It is an open question what happens for the complex contagion model incorporating topological and/or quenched disorder.

\section*{Acknowledgements}
T.H. acknowledges financial support from JSPS (Japan) KAKENHI Grant Numbers JP15K17716, JP16H03939, and JP26310203. T.H. and K.N. acknowledge financial support from JSPS (Japan) KAKENHI Grant Number JP16K05507.

\section*{Author Contribution Statement}

T.H. and K.N. planned the study, derived the results and wrote the manuscript. T.H. performed the numerical simulations.

\end{document}